\documentclass[aps,prd,twocolumn,groupedaddress,amssymb,eqsecnum,showpacs,
nofootinbib]{revtex4}
\usepackage{graphicx}
\usepackage{mathptmx}
\usepackage{bm}
\usepackage{times}

\begin{document}

\preprint{UMHEP-464}

\title{Trans-Planckian signals from the breaking of local Lorentz invariance}

\author{Hael Collins}
\email{hael@physics.umass.edu}
\affiliation{Department of Physics, University of Massachusetts, 
Amherst MA\ \ 01003}
\author{R.~Holman}
\email{rh4a@andrew.cmu.edu}
\affiliation{Department of Physics, Carnegie Mellon University, 
Pittsburgh PA\ \ 15213}

\date{\today}

\begin{abstract}
This article examines how a breakdown of a locally Lorentz invariant, point-like description of nature at tiny space-time intervals would translate into a distinctive set of signals in the primordial power spectrum generated by inflation.  We examine the leading irrelevant operators that are consistent with the spatial translations and rotations of a preferred, isotropically expanding, background.  A few of the resulting corrections to the primordial power spectrum do not have the usual oscillatory factor, which is sometimes taken to be characteristic of a ``trans-Planckian'' signal.  Perhaps more interestingly, one of these leading irrelevant operators exactly reproduces a correction to the power spectrum that occurs in effective descriptions of the state of the field responsible for inflation.
\end{abstract}


\maketitle

\section{Introduction}
\label{introduction}

General relativity postulates that it is always possible to treat the immediate vicinity of any place and time as though it were completely free of the influence of gravity.  Near any point, space and time look flat, regardless of the wider and more complicated environment in which it happens to be situated.  Looking from one place and time to another, the theory tells precisely how these locally flat frames fit together so that the global effects of gravity become apparent.

To all appearances, this postulate seems to be a reasonable principle on which to build a description of our universe.  From terrestrial distances to the size of the observable universe, no discrepancy with the predictions of relativity has been found.  So far, the only somewhat unsettling observation is that during the last five billion years or so, the rate at which the universe is expanding appears to have begun accelerating.  But whether this fact can be attributed to a failure of the theoretical ideas behind relativity or to not having correctly accounted for all of the ingredients of the universe is still unknown.

At the opposite extreme, the idea of a locally flat reference frame is also central for quantum field theory.  How quantum fields propagate through space and how they interact with each other are both strongly constrained if they are assumed to transform consistently with the symmetries of flat space.  This idea additionally influences how to choose an unambiguous and unique lowest energy vacuum state.  And as long as these quantum interactions occur over large enough distances for the influences of gravity to be negligible, it is safe to treat space and time as fixed.

Continuing to still smaller distances, a significant threshold is crossed.  At intervals smaller than this threshold, the standard description of gravity begins to be strongly interacting, if viewed as a quantum theory.  But since gravity corresponds to the dynamics of space-time itself, at such distances it is no longer entirely self-evident that the vicinity of any point can be locally approximated by flat space.  

Many of the attempts thus far to reconcile gravity with the character of a quantum theory have in some way introduced a new length scale, whose role only becomes apparent at very short intervals.  This scale is typically assumed to be close to that same threshold where gravity becomes strongly interacting.  In most of these approaches, space-time at these intervals no longer has the structure and the symmetry of a classical, locally flat background.  

If nature is not locally flat at short distances, some of the usual assumptions about a quantum field theory break down.  The class of allowed interactions could be larger, since how fields interact only needs to be consistent with a reduced set of symmetries.  Moreover, what is the true vacuum state will generally not match with a standard Lorentz-invariant vacuum at these short intervals.  Whether or not such signals of a breakdown of local Lorentz invariance are seen can provide important guidance as to what postulates should be imposed when attempting to incorporate general relativity into a quantum picture of the universe.

Under ordinary circumstances, what happens at such tiny intervals would largely decouple from the interactions and space-time symmetries relevant for any currently accessible experiments, but there is one important exception.  If the universe underwent a stage of accelerated expansion---or {\it inflation\/} \cite{textbooks}---during an early epoch, the ordinary fluctuations of any quantum fields present would be dramatically stretched.  With enough of this stretching, fluctuations that were initially tiny would quickly grow beyond the influence of any {\it subsequent\/} causal process----at least while this stage of accelerated expansion lasts.  Once this phase has ended and the universe starts growing at a decelerating rate, an observer sees farther and farther over time and these fluctuations---until then essentially frozen into the background space-time---can again be seen and come to influence the features of the universe at ever larger scales.  This mechanism for generating a pattern of primordial fluctuations in the background space-time is an essential element of inflation.

If some form of inflation did in fact occur and provided this stage of accelerated expansion lasted sufficiently long, then it should be possible to see any signals of the violation of local Lorentz invariance through characteristic distortions in the pattern of primordial fluctuations.  In practice, these fluctuations are not observed directly, but instead they are seen through their influence on the other ingredients of the universe, appearing eventually among the features of the cosmic microwave background radiation \cite{acoustic,wmap} and the distribution of matter on large scales \cite{sdss}.

As mentioned, one constraint imposed by local Lorentz invariance is on the choice of the vacuum state for the fields present during the inflation.  A fair amount of effort \cite{brandenberger,kaloper,cliff,gary,transplanck,martin,ekp,dan} has already been made to understand how particular departures from this invariance would appear in the microwave background if they arose in new structures in the vacuum state.  Most of this work examined the leading effect, without considering the radiative corrections.  These corrections can actually be quite subtle \cite{einhorn,lowe,fate,taming} when the vacuum departs dramatically from the flat space choice at these short intervals.  To treat them properly requires either imposing very stringent constraints on the behavior of the state at short distances---essentially restricting to just the adiabatic states \cite{emil}---or modifying the propagator to account correctly for the influence of the initial, non-adiabatic state \cite{greens,effective,schalm}.  Despite the fact that many of the previous works have considered models that---in essence---violate local, classical Lorentz invariance in one way or another, none have studied the perhaps simpler problem of determining the influence of symmetry-breaking operators in the effective Lagrangian for inflation.

This article examines how violations of classical Lorentz invariance through such irrelevant symmetry-breaking operators influence the primordial fluctuations produced by inflation.  To isolate the effects of these operators from those produced by departures of the vacuum state from the standard form, we shall choose a conventional vacuum \cite{bunch} throughout---one that matches with the flat-space vacuum over infinitesimal intervals.  So, our goal here is twofold---partially we would like to constrain whether coordinate invariance could be broken at very short distances, but we also wish to learn the extent to which the signatures predicted by choosing non-adiabatic states during inflation can be mimicked by a less radical modification of the theory.

The next section introduces a preferred frame that breaks the symmetry between the spatial and temporal directions and that is appropriate for an inflationary background \cite{mark}.  In this section we also list all of the distinct leading irrelevant operators transforming consistently with this background.  Section \ref{power} then evaluates the effect of each of these operators on the simplest moment of the pattern of primordial fluctuations---its two-point correlator or power spectrum.  In section \ref{models}, we compare these effects with the comparable signals due to non-adiabatic vacuum states.  In some cases we find something familiar but we also find distinctive features too.  Section \ref{conclude} concludes with a brief summary and discussion of our results.

\section{Preliminaries}
\label{prelim}

\subsection{Geometry}

When we consider the possible dynamics of a field theory, it is usually assumed that the underlying symmetry of nature is deeper and more extensive than what is actually realized in the particular background in which it happens to be propagating.  As a simple example, the complete set of invariant quadratic operators that govern the propagation of a scalar field $\varphi$ and are invariant under a general change of coordinates is 
\begin{equation}
{\cal L}_{\rm C} = 
{\textstyle{1\over 2}} g^{\mu\nu}\partial_\mu\varphi \partial_\nu\varphi
- {\textstyle{1\over 2}} \xi R \varphi^2
- {\textstyle{1\over 2}} m^2 \varphi^2 ,
\label{LagrangeC}
\end{equation}
where $R$ is the scalar curvature associated with the metric $g_{\mu\nu}$.  The field $\varphi$ corresponds to the {\it inflaton\/}, the field responsible for the inflationary phase and whose fluctuations, combined with the scalar component of the fluctuations of the metric, results in the primordial perturbations in the background space-time.

Here we examine the signatures of short-distance operators that break this general coordinate invariance, though in a way that is still consistent with the geometry of the background.  ${\cal L}_{\rm C}$ will therefore receive corrections.  These new operators are characterized by whether their importance grows (relevant or marginal corrections) or diminishes (irrelevant corrections) at larger and larger distances.  

Let us begin with an isotropically expanding universe, described by a Robertson-Walker metric, 
\begin{equation}
ds^2 = a^2(\eta)\, \eta_{\mu\nu}\, dx^\mu dx^\nu 
= a^2(\eta) \bigl[ d\eta^2 - d\vec x\cdot d\vec x \bigr] .
\label{metric}
\end{equation}
The rate at which the scale factor $a(\eta)$ changes defines a natural energy scale associated with this geometry,
\begin{equation}
H(\eta) = {a'\over a^2} = {1\over a^2} {da\over d\eta} , 
\label{hubble}
\end{equation}
the Hubble scale.  Except in the case of a de Sitter [$a(\eta)\to -1/(H\eta)$] or a Minkowski [$a(\eta)\to 1$] space-time, this background does not have the maximal, ten-dimensional, possible set of symmetries.  We shall use this metric to define a `preferred frame' for our theory, one where the space-time is organized into spatial slices orthogonal to the vector, 
\begin{equation}
n_\mu = \bigl( a(\eta), 0, 0, 0 \bigr) .
\label{ndef}
\end{equation}
Because spatial symmetries---translations and rotations---are preserved by this frame, we shall allow only those operators that remain invariant under this smaller set of symmetries.

By removing the components of the metric that lie along the same direction, the normal defines an induced metric for the spatial surfaces orthogonal to it, 
\begin{equation}
h_{\mu\nu} = g_{\mu\nu} - n_\mu n_\nu .
\label{induced}
\end{equation}
In our Robertson-Walker frame, this induced metric is flat,
\begin{equation}
h_{\mu\nu}\, dx^\mu dx^\nu = - a^2(\eta)\, d\vec x\cdot d\vec x , 
\label{RWinduced}
\end{equation}
being only rescaled from one surface to the next through the appearance of the scale factor.

We can use each of these tensors to construct two more by projecting the derivative of $n_\mu$ onto the normal and the transverse directions, 
\begin{equation}
n^\lambda \nabla_\lambda n_\mu 
\label{nDn}
\end{equation}
and 
\begin{equation}
K_{\mu\nu} = h_\mu^{\ \ \lambda} \nabla_\lambda n_\nu .
\label{extrinsic}
\end{equation}
This latter tensor is the standard extrinsic curvature.  In the Robertson-Walker frame, the first vanishes while second is proportional to the Hubble scale, 
\begin{equation}
K_{\mu\nu}\, dx^\mu dx^\nu = - a^2 H\, d\vec x\cdot d\vec x .  
\label{RWextrinsic}
\end{equation}
Together with the standard covariant tensors,
\begin{equation}
g_{\mu\nu}, \nabla_\mu, R_{\lambda\mu\nu\sigma}, \ldots
\label{covten}
\end{equation}
we shall use these additional objects,
\begin{equation}
n_\mu, h_{\mu\nu}, K_{\mu\nu}, 
\label{nocovten}
\end{equation}
to generate the corrections to the free field theory described by ${\cal L}_{\rm C}$.

One final ingredient we shall use is a non-analytic operator ${\cal D}$ that essentially extracts a factor of the magnitude of the spatial momentum, 
\begin{equation}
{\cal D} \equiv 
\bigl( h^{\mu\nu}\nabla_\mu\nabla_\nu - K n^\mu\nabla_\mu \bigr)^{1/2} ; 
\label{Dopdef}
\end{equation}
despite its somewhat complicated form, ${\cal D}$ becomes more familiar once we have written it in the Robertson-Walker frame, 
\begin{equation}
{\cal D} = {1\over a} 
\bigl( - \vec\nabla\cdot\vec\nabla \bigr)^{1/2} . 
\label{RWDop}
\end{equation}
Thus, for example, acting with ${\cal D}$ on the scalar field, written in its operator expansion,
\begin{equation}
\varphi(\eta,\vec x) = 
\int {d^3\vec k\over (2\pi)^3}\, \bigl[ 
U_k(\eta) e^{i\vec k\cdot \vec x} a_{\vec k} 
+ U_k^*(\eta) e^{-i\vec k\cdot \vec x} a^\dagger_{\vec k} \bigr] , 
\label{phiop}
\end{equation}
yields 
\begin{equation}
{\cal D}\varphi(\eta,\vec x) = 
\int {d^3\vec k\over (2\pi)^3}\, {k\over a}\bigl[ 
U_k(\eta) e^{i\vec k\cdot \vec x} a_{\vec k} 
+ U_k^*(\eta) e^{-i\vec k\cdot \vec x} a^\dagger_{\vec k} \bigr] . 
\label{Dphiop}
\end{equation}

\subsection{Symmetry-breaking operators}

It is now only a matter of constructing all the independent operators that can be assembled from the elements just described to learn how the signals of broken covariance would appear.  Most often, the focus is on the relevant or marginal operators, since they grow most prominent at low energies, where we have the most direct experience.  While we shall include such terms within our catalogue of operators here, our emphasis will instead be on the leading irrelevant operators, since they are the ones that best imitate the trans-Planckian signatures that are generated by a non-adiabatic vacuum during inflation.

In a general, maximally asymmetric background, the number of distinct symmetry-breaking operators at any particular order can be quite large.  In the Robertson-Walker background, however, only a few of the symmetries are broken.  Its main feature is that it treats the temporal and spatial directions differently, so the operators can be largely characterized by their relative numbers of time and space derivatives.  The former scale maximally as some power of $H$, the Hubble scale, while the latter scale as powers of the spatial momentum, $\vec k$.

The simplest class of corrections to modify the power spectrum are those that are quadratic in the field, $\varphi$, so we shall discuss operators with this structure.  Starting at dimension three---the only dimension-two operator being just the mass term---we have two possibilities:  the operator,
\begin{equation}
{\textstyle{1\over 3}} K \varphi^2 ,
\label{dim3a}
\end{equation}
and the non-analytic operator, 
\begin{equation}
\varphi{\cal D}\varphi . 
\label{dim3b}
\end{equation}
These two are the only new relevant operators for this background.  

At the next order, the four independent dimension-four terms are 
\begin{equation}
{\textstyle{1\over 9}} K^2\varphi^2,\
{\textstyle{1\over 3}} K\varphi{\cal D}\varphi,\
- h^{\mu\nu}\nabla_\mu\varphi\nabla_\nu\varphi,
\label{dim4}
\end{equation}
beyond the standard kinetic and conformal terms which appeared already in ${\cal L}_{\rm C}$.  Together these five operators describe all the possible renormalizable corrections to the covariant Lagrangian, 
\begin{eqnarray}
{\cal L}_{\rm R} &\!\!\!=\!\!\!& 
{\textstyle{1\over 3}} c_1 M K \varphi^2 
+ c_2 M\varphi{\cal D}\varphi 
\nonumber \\
&&
+ {\textstyle{1\over 9}} c_3 K^2\varphi^2 
+ {\textstyle{1\over 3}} c_4 K\varphi{\cal D}\varphi 
- c_5 h^{\mu\nu}\nabla_\mu\varphi\nabla_\nu\varphi . 
\qquad
\label{LR}
\end{eqnarray}
Here we have introduced a new mass scale $M$ associated with whatever dynamics or principle is responsible for the broken symmetry.  In the Robertson-Walker frame, ${\cal L}_{\rm R}$ becomes 
\begin{eqnarray}
{\cal L}_{\rm R} &\!\!\!=\!\!\!& 
c_1\, MH \varphi^2 
+ {c_2\over a}\, M\varphi\bigl( -\vec\nabla\cdot\vec\nabla \bigr)^{1/2}\varphi 
\nonumber \\
&&
+ c_3\, H^2\varphi^2
+ {c_4\over a}\, H \varphi\bigl( -\vec\nabla\cdot\vec\nabla \bigr)^{1/2}\varphi 
+ {c_5\over a^2}\, \vec\nabla\varphi\cdot\vec\nabla\varphi . 
\qquad\ 
\label{RWLR}
\end{eqnarray}

Our main interest here is the set of leading---dimension-five---irrelevant operators.  All of the standard covariant terms must contain an even power of derivatives; so at this order, the only possible operators are those explicitly violating the coordinate invariance.  The many ways of contracting the many indices, combined with the choices for how the derivatives act on the fields or on the background, means that the number of operators proliferates very rapidly at higher orders.  But in a fairly symmetric background, such as the Robertson-Walker space-time, only a small number of these produce distinct corrections.  Moreover, in an inflationary setting, the Hubble scale typically changes only slowly, $H^2 \gg H'$, so among the terms where derivatives act on the background, those scaling as $H$ to some power produces the dominant effects.

Based upon these observations, we select four of the dimension-five operators that are quadratic in the field and that essentially capture all of distinctive scalings possible,
\begin{eqnarray}
{\cal L}_{\rm NR} 
&\!\!\!=\!\!\!& {d_1\over 27M} K^3\varphi^2 
+ {d_2\over 9M} K^2\varphi{\cal D}\varphi 
\nonumber \\
&& 
- {d_3\over 3M} K h^{\mu\nu}\nabla_\mu\varphi\nabla_\nu\varphi 
+ {d_4\over M} \varphi{\cal D}^3\varphi , 
\label{LNR}
\end{eqnarray}
which reduces to
\begin{eqnarray}
{\cal L}_{\rm NR} 
&\!\!\!=\!\!\!& {d_1\over M} H^3 \varphi^2 
+ {d_2\over aM} H^2 \varphi\bigl( -\vec\nabla\cdot\vec\nabla \bigr)^{1/2}\varphi 
\nonumber \\
&& 
+ {d_3\over a^2M} H \vec\nabla\varphi\cdot\vec\nabla\varphi 
+ {d_4\over a^3M} \varphi\bigl( -\vec\nabla\cdot\vec\nabla \bigr)^{3/2}\varphi ,
\qquad 
\label{RWLNR}
\end{eqnarray}
in the Robertson-Walker frame.

\section{Trans-Planckian corrections to the power spectrum}
\label{power}

The symmetry-breaking terms can have a small effect on the pattern of primordial fluctuations which in their turn influence the cosmic microwave background and the formation of structures in the universe.  To extract the basic signals of these effects, we calculate the corrections to the power spectrum of the scalar field due to the nonrenormalizable terms, the leading representative set of which composes ${\cal L}_{\rm NR}$.  We shall work in the de Sitter limit, which is the simplest to treat analytically, although our operators are not in fact invariant under all of the generators of the symmetry group of de Sitter space.

The actual pattern of primordial perturbations imprinted on the background can be characterized by how fluctuations at different places happen to be correlated with each other.  For the scalar fluctuations described here, such correlations are captured by the expectation value of some number of fields each evaluated at an arbitrary position, 
\begin{equation}
\langle 0(\eta)|\, \varphi(\eta,\vec x_1) \varphi(\eta,\vec x_2) \cdots 
\varphi(\eta,\vec x_n)|\, 0(\eta)\rangle . 
\label{npoint}
\end{equation}
In practice, the independent effects of higher order moments on the cosmic microwave radiation and the distribution of structures over large scales is extremely small, so observationally the most important moment is just the correlator between two points,
\begin{equation}
\langle 0(\eta)|\, \varphi(\eta,\vec x)\varphi(\eta,\vec y)|\, 0(\eta)\rangle . 
\label{twopointeg}
\end{equation}

The power spectrum corresponds to the Fourier transform of this two-point function, usually normalized with some of its momentum dependence extracted, 
\begin{eqnarray}
&&\!\!\!\!\!\!\!\!\!\!\!\!\!\!\!\!\!\!\!\!\!\!\!\!\!\!\!\!\!\!
\langle 0(\eta)|\, \varphi(\eta,\vec x) \varphi(\eta,\vec y)|\, 0(\eta)\rangle 
\nonumber \\
&& = \int {d^3\vec k\over (2\pi)^3}\, e^{i\vec k\cdot(\vec x-\vec y)} 
\biggl[ {2\pi^2\over k^3} P_k(\eta) \biggr] . 
\label{powerdef}
\end{eqnarray}
Notice that in writing the two-point function in this form---with some time-dependence in the state---we have implicitly assumed that the time-evolution of the components of a matrix element is determined by the interaction picture, where the evolution of the field is generated by the free Lagrangian, ${\cal L}_{\rm C}$, and that of the state is generated by the interacting parts, here ${\cal L}_{\rm NR}$.

Our purpose is to extract the general signatures in the power spectrum from the symmetry breaking operators, specifically how they might resemble or differ from the signatures of trans-Planckian structures in the initial state of the inflaton.  To make these differences especially clear, we shall evaluate the power spectrum in a purely Bunch-Davies state.  With the possibility of dimension-five operators, the leading corrections from the irrelevant operators still scale as $H/M$, which is the typical size for the signals of trans-Planckian physics.  However, we shall also find that some of the corrections scale as $k\eta_0$---where $\eta_0$ is the time at which the inflationary expansion begins.  In the effective theory treatments of the trans-Planckian problem \cite{greens,effective,schalm}, similar $k$-dependent corrections were also found, although the analogue of $\eta_0$ that appears there is only a cutoff for the effective theory and not the physical beginning of inflation, as it is here.  The terms scaling as $k\eta_0$ inevitably diverge if the duration of the inflationary period is taken to be arbitrarily long:  $\eta_0\to -\infty$ in conformal coordinates.

Sometimes, rather than referring to the time at which the inflationary expansion began, we shall introduce a `pivot point' value for the spatial momenta, $k_*$.  We define this $k_*$ to correspond to a mode whose physical size is equal to our new effective theory scale $M$ just as inflation begins,
\begin{equation}
{k_*\over |a(\eta_0)|} \equiv M . 
\label{kstardef}
\end{equation}
If we think of $M$ as the Planck mass---although it need not be---then $k_*$ indicates the threshold at which the `trans-Planckian' modes begin.  In the limit $\eta_0\to -\infty$, $k_*\to 0$, which simply means that all the modes we see today were at one time trans-Planckian.

\subsection{de Sitter space}

As mentioned, we evaluate the corrections to the power spectrum in the limit of a pure de Sitter space-time.  In de Sitter space, the energy density of the vacuum remains constant and so the rate of expansion is also a constant, $H(\eta)\to H$.  de Sitter space is also one of the three maximally symmetric space-times, so the operators that we introduced in the previous section explicitly break some of the symmetries of the background.  The scale factor in this case becomes 
\begin{equation}
a(\eta) = - {1\over H\eta} , 
\label{dSscale}
\end{equation}
which is chosen so to agree with our previous stated convention that $\eta\to -\infty$ indicates the far past; the infinitely far future corresponds then to $\eta\to 0$. 

Since the energy density is everywhere constant, the curvature of de Sitter space is constant too, $R = 12H^2$, so that there is no real distinction between the mass term and the conformal-coupling term in the free Lagrangian, ${\cal L}_{\rm C}$, and so we set $\xi=0$.  A free field in this background then satisfies a simple Klein-Gordon equation, 
\begin{equation}
\biggl[ {\partial^2\over\partial\eta^2} 
- {2\over\eta}{\partial\over\partial\eta} 
- \vec\nabla\cdot\vec\nabla 
+ {1\over\eta^2} {m^2\over H^2} \biggr] \varphi = 0 , 
\label{KG}
\end{equation}
which correspondingly implies a differential equation for the mode functions, \begin{equation}
U^{\prime\prime}_k - {2\over\eta} U'_k
+ \biggl[ k^2 + {1\over\eta^2} {m^2\over H^2} \biggr] U_k = 0 , 
\label{KGk}
\end{equation}
where the $U_k(\eta)$ are the eigenmodes associated with the operator expansion of the field,
\begin{equation}
\varphi(\eta,\vec x) 
= \int {d^3\vec k\over (2\pi)^3}\, 
\bigl[ U_k(\eta) e^{i\vec k\cdot\vec x} a_{\vec k} 
+ U_k^*(\eta) e^{-i\vec k\cdot\vec x} a_{\vec k}^\dagger \bigr] . 
\label{opexand}
\end{equation}
If we rescale the mode functions with a suitable factor of the conformal time, $U_k(\eta) = \eta^{3/2}Z_\nu(k\eta)$, and define a dimensionless variable $z=k\eta$, then the Klein-Gordon equation for the modes assumes the the form of Bessel's equation, 
\begin{equation}
{d^2Z_\nu\over dz^2} + {1\over z} {dZ_\nu\over dz} 
+ \biggl( 1 - {\nu^2\over z^2} \biggr) Z_\nu = 0 , 
\label{bessel}
\end{equation}
where 
\begin{equation}
\nu = \sqrt{ {9\over 4} - {m^2\over H^2} } . 
\label{nudef}
\end{equation}
The normalization of the mode $U_k(\eta)$ is entirely fixed by the equal time commutation relation between the field $\varphi$ and its conjugate momentum, but the second constant of integration is determined by the choice of the state.  The standard choice is the Bunch-Davies state \cite{bunch}, $|0\rangle$, which matches with the form of the Minkowski vacuum at short distances and is functionally 
\begin{equation}
U_k(\eta) = {\sqrt{\pi}\over 2} H \eta^{3/2} H_\nu^{(2)}(k\eta) , 
\label{vacuummodes}
\end{equation}
where the $H_\nu^{(2)}(k\eta)$ is a Hankel function of the second type. 

In an inflating universe, we can make one final simplifying approximation since the effective mass of the scalar field must be quite small compared with the Hubble scale, $m\ll H$.  Therefore we calculate the power spectrum in the limit of a massless field, where $\nu={3\over 2}$; the only danger in doing so is that the strictly massless theory can introduce infrared divergences which are an artifact of setting $m\to 0$ and which can be removed by taking a small but finite value for the mass of the scalar field.  In the massless limit, the Bunch-Davies mode functions simplify yet further to 
\begin{equation}
U_k(\eta) = {H\over k\sqrt{2k}} (i-k\eta) e^{-ik\eta} .  
\label{mmcmodes}
\end{equation}

To have a point of comparison for the corrections from the symmetry-breaking operators, let us calculate the power spectrum of this simplest of settings,
\begin{equation}
P_k(\eta) 
= {k^3\over 2\pi^2} U_k(\eta) U_k^*(\eta) 
= {H^2\over 4\pi^2} (1 + k^2\eta^2) . 
\label{BDpower}
\end{equation}
The physically interesting modes---those that have been stretched well outside the horizon during inflation to become a sort of noise frozen into the background space-time---correspond to those where $k\eta\to 0$.  For these modes the power spectrum is essentially flat.

\subsection{Corrections}

Although the symmetry-breaking terms are also quadratic in the field, we shall assume that their effect is small so that they can be treated as perturbations.  Since we have no knowledge of how long a stage of inflationary expansion might have lasted or what might have preceded it,\footnote{Even were we to assume an epoch of inflation extending arbitrarily far into the past, an $S$-matrix description---just as for a purely de Sitter background \cite{witten}---would not be appropriate.} we apply the Schwinger-Keldysh \cite{sk} approach for evaluating the corrections to the two-point function.\footnote{A description of the Schwinger-Keldysh approach as it is applied to an inflationary setting is given in \cite{greens} and in \cite{weinberg}.}  The Schwinger-Keldysh formalism evolves both the state $|0\rangle$ and its dual $\langle 0|$ from an initial configuration at $\eta_0$ to an arbitrary later time $\eta$, 
\begin{equation}
\langle 0(\eta)|\, \varphi(\eta,\vec x)\varphi(\eta,\vec y)|\, 0(\eta)\rangle ,
\label{twopoint}
\end{equation}
where the time-evolution of the state is given in the interaction picture by 
\begin{equation}
|0(\eta)\rangle = Te^{-i\int_{\eta_0}^\eta d\eta'\, H_I(\eta')}\, |\, 0\rangle .
\label{dyson}
\end{equation}
Here we have written the initial state more succinctly as $|0(\eta_0)\rangle = |0\rangle$.  $H_I$ is the interaction Hamiltonian, which is, considering only the irrelevant symmetry-breaking operators, 
\begin{equation}
H_I(\eta) = - \int d^3\vec x\, \sqrt{-g}\, {\cal L}_{\rm NR} . 
\label{IHam}
\end{equation}

Before evaluating the power spectrum to first order in the corrections, we should first compare the initial and final times used for the time-evolution of the state with the scales that are important for the later cosmology.  For these modes, $k$ is very small compared with the conformal time by the end of inflation, $k\eta\to 0$.  Therefore, we shall neglect terms that vanish in this limit.  Since this is an inherently long-distance limit, we shall occasionally meet with mild divergences arising because we have neglected the mass of the field, a property of the theory that also obviously persists to long distances.

Furthermore, the modes responsible for the structures we are observing today should have been well within the horizon at the beginning of inflation, $|k\eta_0| \gg 1$.  We shall therefore often take the limit where $k\eta_0\to -\infty$, neglecting terms that are small in this limit.  In terms of the pivot momentum $k_*$ that we defined earlier, in a de Sitter background it is defined through 
\begin{equation}
\eta_0 = - {1\over k_*} {M\over H} . 
\label{dSkstar}
\end{equation}
As we shall see, the effects that scale with a sufficient power of the spatial momentum are especially sensitive to when the initial time is chosen.

Having established these preliminaries, we can evaluate the leading corrections to the power spectrum from the dimension-five symmetry-breaking operators, listed in ${\cal L}_{\rm NR}$ in Eq.~(\ref{RWLNR}), to obtain
\begin{eqnarray}
P_k(\eta) 
&\!\!\!=\!\!\!& {H^2\over 4\pi^2} \biggl[ 1 + k^2\eta^2 
- 2 {H\over M} \bigl[ 
d_1\, {\cal I}_4(k\eta,k\eta_0) 
\nonumber \\
&&\qquad
-\, d_2\, {\cal I}_3(k\eta,k\eta_0) 
- d_3\, {\cal I}_2(k\eta,k\eta_0) 
\nonumber \\
&&\qquad
-\, d_4\, {\cal I}_1(k\eta,k\eta_0) 
\bigr] + \cdots \biggr] , 
\label{LNRpower}
\end{eqnarray}
where we have treated the corrections as small effects.  The function ${\cal I}_n(z,z_0)$ that appears in this expression corresponds to the following set of dimensionless integrals,
\begin{eqnarray}
{\cal I}_n(z,z_0)
&\!\!\!=\!\!\!& \int_{z_0}^z {dz'\over z^{\prime n}}\, \Bigl[ 
\bigl[ 1 - z^2 + 4zz' - z^{\prime 2} + z^2z^{\prime 2} \bigr] \sin\bigl[ 2(z-z')\bigr] 
\nonumber \\
&&\qquad\quad 
-\, 2(z-z')[1+zz'] \cos\bigl[ 2(z-z')\bigr] \Bigr] . 
\label{Iintdef}
\end{eqnarray}

At a first glance, and as expected, all of the new corrections are suppressed by $H/M$, as is familiar from a variety models that include some non-standard, short-distance structure in the inflaton's state \cite{brandenberger,kaloper,cliff,gary,transplanck,martin,ekp,dan}.  However, this is not the only dimensionless scale available.  When we extract the asymptotic behavior of these integrals in the 
\begin{equation}
k\eta\to 0\ , \qquad k\eta_0\to -\infty 
\label{limits}
\end{equation}
limits, we shall find that several of the corrections also depend sensitively on $k\eta_0$.  Applying these limits, we look at the four corrections one by one.

\subsubsection{The correction from $K^3\varphi^2 \to H^3 \varphi^2$}

The first of the corrections, which contains only time-derivatives, produces a small correction to the power spectrum,
\begin{equation}
P_k(\eta) = {H^2\over 4\pi^2} \biggl[ 1 
+ {4\over 3}d_1\, {H\over M} 
\bigl[ \ln|2k\eta| - 2 + \gamma \bigr]
+ \cdots \biggr] . 
\label{LNRpower1}
\end{equation}
The new terms are all accompanied by the standard small factor of $H/M$, though there are already, even in this fairly innocuous term, a few differences from more standard trans-Planckian corrections.  First, the correction contains a mild logarithmic divergence, $\ln|2k\eta|$.  This divergence occurs only in the long-distance, $k\eta\to 0$ limit, but its origin is quite simple to understand.  In a pure de Sitter space-time, $H$ is constant so the interaction between the field and the background given by $H^3\varphi^2$ is itself essentially a mass term.  If we take a very small ($m\ll H$), but finite mass for the field, then leading contribution to the power spectrum in the $k\eta\to 0$ limit scales as 
\begin{equation}
P_k(\eta) = {H^2\over 4\pi^2} {4^\nu\, \Gamma^2(\nu)\over 2\pi} 
|k\eta|^{3-2\nu} + \cdots , 
\label{nonminimal}
\end{equation}
where $\nu$ is given in Eq.~(\ref{nudef}).  Expanding near $\nu\sim {3\over 2}$ yields exactly the same structure as this ``trans-Planckian'' correction,
\begin{equation}
P_k(\eta) = {H^2\over 4\pi^2} \biggl[ 1 
+ {2\over 3} {m^2\over H^2} 
\bigl[ \ln|2k\eta| - 2 + \gamma \bigr]
+ \cdots \biggr] , 
\label{powermass}
\end{equation}
if we replace 
\begin{equation}
m^2 \to 2d_1\, {H^3\over M} .
\label{massequiv}
\end{equation}
In the more realistic setting of a slowly rolling period of inflation, however, $H$ does contain some time dependence, so the effect of this term no longer is equivalent to that of a simple mass term.

The second difference, which appears in the next correction as well, is that the $H/M$ is not accompanied by a modulating factor, such as usually occurs in trans-Planckian corrections.  Very typically, imposing some cut-off or some modification in the dispersion relation of the inflaton introduces a `ringing' in the power spectrum.  This `ringing' appears as an oscillatory factor, such as $\cos(2M/H)$.  The correlation between the amplitude of the correction and the frequency of the modulation is often taken as a distinctive sign of a trans-Planckian effect.  Of course, such a correlation can still be taken as a distinctive signature of a {\it state\/} that incorporates some trans-Planckian structure, as opposed to a theory where some symmetries are explicitly broken in the Lagrangian in the trans-Planckian regime.  But in other instances, as we shall soon see, this clear ability to distinguish the source of a trans-Planckian effect breaks down.

\subsubsection{The correction from $K^2\varphi{\cal D}\varphi \to H^2 \varphi\bigl( -\vec\nabla\cdot\vec\nabla \bigr)^{1/2}\varphi$}

It might be thought that any operator that contains a factor of the spatial momentum would inevitably give corrections that diverge as $k$ grows too large.  However, the correction from the next operator, $H^2 \varphi\bigl( -\vec\nabla\cdot\vec\nabla \bigr)^{1/2}\varphi$, shows that this fear is not realized,
\begin{equation}
P_k(\eta) = {H^2\over 4\pi^2} \biggl[ 1 
+ d_2\, {H\over M} 
\biggl[ \pi + {\cos(2k\eta_0)\over k\eta_0}
\biggr]
+ \cdots \biggr] . 
\label{LNRpower2}
\end{equation}
As with the previous correction, the observable effect of this operator is small---unless the inflationary stage is very short, so that some of the modes are near the $k\eta_0=-1$ limit---with only a mild scale dependence that is implicit in $H$ when we leave the ideal realm of de Sitter space and return to a slowly rolling space-time.  Also, the modulating factor is again absent in the leading effect.  Note that in a pure de Sitter space-time, where $H$ is constant, this correction is largely unobservable since it describes only a small rescaling.

\subsubsection{The correction from $K h^{\mu\nu}\nabla_\mu\varphi\nabla_\nu\varphi \to H \vec\nabla\varphi\cdot\vec\nabla\varphi$}

The first appearance of a direct sensitivity on the wave number occurs in the next term,
\begin{equation}
P_k(\eta) = {H^2\over 4\pi^2} \biggl[ 1 
+ d_3\, {H\over M} \bigl[ 3 + \cos(2k\eta_0) \bigr] 
+ \cdots \biggr] , 
\label{LNRpower3}
\end{equation}
or in terms of the threshold momentum, $k_*$, 
\begin{equation}
P_k(\eta) = {H^2\over 4\pi^2} \biggl[ 1 
+ d_3\, {H\over M} \biggl[ 3 + \cos\biggl( 2{k\over k_*} {M\over H} \biggr) \biggr] 
+ \cdots \biggr] . 
\label{LNRpower3ks}
\end{equation}
In this form, the effect of the initial time is rather benign, since it only appears in the argument of the cosine factor.  If the modes that we observe today were much smaller than the scale $1/M$ at the beginning of inflation, which corresponds to $k/k_*$ being extremely large, such a term would introduce some fundamental noise into the power spectrum since we would not be able to resolve the frequency of the modulation, though its amplitude ($H/M$) would still be small.

\subsubsection{The correction from $\varphi{\cal D}^3\varphi \to \varphi\bigl( -\vec\nabla\cdot\vec\nabla \bigr)^{3/2}\varphi$}

We come now to the last and most interesting of the new corrections, that which contains the maximal number of spatial derivatives at this order.  Its correction does depend sensitively on the initial time, not only through an oscillatory term, but more importantly through its amplitude,
\begin{equation}
P_k(\eta) = {H^2\over 4\pi^2} \biggl[ 
1 + d_4\, {H\over M} k\eta_0 \cos(2k\eta_0) + \cdots 
\biggr] 
\label{LNRpower4}
\end{equation}
or equivalently
\begin{equation}
P_k(\eta) = {H^2\over 4\pi^2} \biggl[ 
1 - d_4\, {k\over k_*} \cos\biggl( 2{k\over k_*} {M\over H} \biggr) + \cdots 
\biggr] . 
\label{LNRpower4ks}
\end{equation}
Since this correction depends linearly on the wave number $k$, it can overwhelm the standard prediction once $k>k_*$.  Remember that $k_*$ is the wave number of a mode that had a wavelength of $1/M$ at the beginning of inflation---it is the threshold between ordinary and trans-Planckian modes when $M=M_{\rm pl}$.

If such a symmetry-breaking operator is present in the theory, then it allows only a narrow window of modes that can be responsible for the features that we see in the cosmic microwave background.  The widest allowed range,
\begin{equation}
H < {k\over|a(\eta_0)|} < M , 
\label{scalelimits}
\end{equation}
corresponds to when the minimal amount of inflation occurs---that is, when a fluctuation of the order of the Hubble horizon at the beginning of inflation was stretched just enough to encompass the observed universe today.  The upper bound is fixed and is imposed by the requirement that the corrections to the power spectrum, coming from an operator such as $\varphi{\cal D}^3\varphi$, should remain perturbative.  In de Sitter space, these bounds can also be written as 
\begin{equation}
{H\over M} < {k\over k_*} < 1 .
\label{modelimits}
\end{equation}
Having more than this minimal amount of inflation further constricts this range---the upper bound remains fixed, but the lower bound increases since the largest observable modes in the microwave background would have been well within the Hubble horizon even at the beginning of inflation.

We might worry that in the case of `just enough' inflation---where a fluctuation the size of the horizon at the beginning of inflation is just re-entering the horizon today---some of the observable modes $k/k_*$ can be quite small, of the order of $H/M$.  In this case we cannot assume that $k\eta_0\to -\infty$, as we have done.  However, the largest allowed modes at the beginning of inflation can never have $|k\eta_0|$ smaller than one.  Even in the limiting case, $k\eta_0=-1$, all of the integrals ${\cal I}_n(z,z_0)\sim {\cal O}(1)$ (for $n=1,2,3,4$) up to small, order ${\cal O}(z^2)$ corrections.

\subsection{Higher order operators}

Among the dimension-five operators that we have analyzed, the one with the most dramatic potential signal is the operator $\varphi{\cal D}^3\varphi$.  As we shall see in the next section, its signal is of the same form as one produced in an effective-state treatment of the trans-Planckian problem.  This operator is admittedly of a rather peculiar form, since it contains the non-analytic derivative operator ${\cal D}$, defined in Eq.~(\ref{Dopdef}).  However, none of the interesting effects that it produces are unique to ${\cal D}$ and we find many examples of similar effects on the power spectrum produced by higher dimensional operators.

As an example, let us consider the following dimension-six operator, 
\begin{equation}
{d_5\over M^2} {(\vec\nabla\cdot\vec\nabla\varphi)^2\over a^4} , 
\label{dim6eg}
\end{equation}
which we have already written for a Robertson-Walker frame.  Its contribution to the power spectrum, again evaluated in the $k\eta\to 0$ and $k\eta_0\to -\infty$ limits, is 
\begin{equation}
P_k(\eta) = {H^2\over 4\pi^2} \biggl[ 1 
- d_5 {H^2\over M^2} (k\eta_0)^2 \cos(2k\eta_0) 
+ \cdots \biggr] 
\label{LNR6power}
\end{equation}
or
\begin{equation}
P_k(\eta) = {H^2\over 4\pi^2} \biggl[ 1 
- d_5 {k^2\over k_*^2} \cos\biggl( 2 {k\over k_*} {M\over H} \biggr) 
+ \cdots \biggr] , 
\label{LNR6powerk}
\end{equation}
in terms of the threshold wave number $k_*$.

So the strict constraint---either on the existence of operators with the maximal number of spatial derivatives at each order in the effective theory or on the duration of inflation---does not depend on whether the theory contains non-analytic structures such as ${\cal D}$.  At $k\sim k_*$, all operators of the general form
\begin{equation}
{1\over M^{2n}} {1\over a^{2n+2}} (\vec\nabla\cdot\vec\nabla)^{n+1} \varphi^2 
\qquad n=1,2,3,\ldots , 
\label{dimN}
\end{equation}
contribute equally to the power spectrum and so the theory no longer admits a perturbative description of processes.

\subsection{Lower order operators}

Although our interest has been primarily in the irrelevant operators that break local Lorentz invariance, the relevant operators can produce, in principle, a much larger effect on the power spectrum and are therefore much more strongly constrain the amount of symmetry breaking that could have occurred at long distances during an inflationary era.  For example, the two dimension-three operators considered earlier, 
\begin{eqnarray}
{\cal L}_R
&\!\!\!=\!\!\!& 
{1\over 3} c_1\, MK\varphi^2 + c_2\, M\varphi{\cal D}\varphi
\nonumber \\
&\!\!\!=\!\!\!& 
c_1\, MH\varphi^2 
+ c_2\, {M\over a}\varphi \bigl( - \vec\nabla\cdot\vec\nabla \bigr)^{1/2} \varphi , 
\label{LRoperators}
\end{eqnarray}
produce the following effects in the power spectrum, again in the de Sitter limit with a massless, minimally coupled field,
\begin{eqnarray}
P_k(\eta) 
&\!\!\!=\!\!\!& {H^2\over 4\pi^2} \biggl[ 1 + k^2\eta^2 
+ 2\, c_1 {M\over H} \, {\cal I}_4(k\eta,k\eta_0) 
\nonumber \\
&&\qquad
+ 2\, c_2 {M\over H} \, {\cal I}_3(k\eta,k\eta_0) 
+ \cdots 
\bigr]\biggr] . 
\label{LRpower}
\end{eqnarray}
For the physically relevant modes ($k\eta\to 0$ and $k\eta_0\to -\infty$) we find effects with essentially the same behavior as before, 
\begin{eqnarray}
P_k(\eta) 
&\!\!\!=\!\!\!& {H^2\over 4\pi^2} \biggl[ 1 
- {4\over 3} \, c_1 {M\over H} \, \bigl[ 
\ln|2k\eta| - 2 + \gamma \bigr]
\nonumber \\
&&\qquad\quad 
+ c_2 {M\over H} \, \biggl[ \pi + {\cos(2k\eta_0)\over k\eta_0}
\biggr] 
+ \cdots 
\biggr] , \qquad
\label{LR3power}
\end{eqnarray}
except that, whereas an $H/M$ suppression occurred before, here the signals are enhanced by $M/H$ and are therefore much more strongly constrained by observations.

\section{Prior models}
\label{models}

Taken by themselves, symmetry breaking operators can provide an additional source for trans-Planckian effects, but to what extent can they be distinguished from other mechanisms for producing such effects?  A broad class of models for treating possible trans-Planckian signatures can be characterized by the fact that some new principle or property of nature---which becomes important only at very short distances---modifies the choice for the vacuum state.  In these models, the correct vacuum state is not the sort of adiabatic vacuum state that we would have anticipated by extrapolating our understanding of nature at large scales to such arbitrarily small scales.  

Within these models with modified choices for the vacuum state, there are two further classes---one where a particular assumption is made about what new thing happens at distances approaching the Planck scale and another that treats the states using an effective theory description.  While the exact details depend on the specific model, a typical prediction from this first class is that the power spectrum receives a oscillating correction whose amplitude, $H/M$, and frequency, $M/H$, are correlated.  For example, if we take the de Sitter limit of the prediction of \cite{martin}, we obtain,
\begin{equation}
P_k(\eta) = {H^2\over 4\pi^2} \biggl\{ 
1 - 2 {\cal O}(1) {H\over M} \cos\biggl[ 2{M\over H} + \phi \biggr] \biggr\} , 
\label{Pkmartin}
\end{equation}
where `${\cal O}(1)$' is a model-dependent, order one parameter and $\phi$ is an arbitrary phase.

Such a signal can in principle be distinguished from the $H/M$ effects produced by the symmetry-breaking operators here.  The corrections from the first two, 
\begin{equation}
H^3\varphi^2, \quad H^2\varphi \bigl( - \vec\nabla\cdot\vec\nabla \bigr)^{1/2} \varphi , 
\label{firsttwo}
\end{equation}
are not accompanied by any oscillatory factors and the correction from the third operator, 
\begin{equation}
H \vec\nabla\varphi\cdot\vec\nabla\varphi ,
\label{thirdone}
\end{equation}
oscillates much more rapidly since its phase is directly proportional to $k$ rather than the weaker $k$-dependence inherited through $H$, as in Eq.~(\ref{Pkmartin}), once we return to a true slowly rolling model.

It is much more interesting when we compare with some of the signatures from the second class---effective theory modifications of the vacuum state.  In addition to corrections that scale as $H/M$ as above, these theories also produce corrections that scale as $k/k_*$.  As an example, and again up to dimensionless, model-dependent ${\cal O}(1)$ coefficients, a fairly standard prediction for the power spectrum from this effective state approach is \cite{ekp,schalm} 
\begin{equation}
P_k(\eta) = {H^2\over 4\pi^2} \biggl\{ 
1 + {\cal O}(1) {k\over k_*} \sin\biggl[ 2 {k\over k_*} {M\over H} \biggr] \biggr\} . 
\label{Pkschalm}
\end{equation}
This signal is essentially identical to that produced by the dimension-five operator $\varphi{\cal D}^3\varphi$, although the notation here does conceal a formal difference.  Unlike the $\eta_0$ used to define $k_*$ earlier, the $\eta_0$ implicit in Eq.~(\ref{Pkschalm}) is not necessarily the true beginning of the inflationary expansion, but rather defines an initial spatial surface on which the effective state is defined.

Perhaps it should not be entirely surprising that these two approaches have yielded similar predictions.  Once we include new structures in the effective state at short distances, the state itself can break the same space-time symmetries as an operator such as $\varphi{\cal D}^3\varphi$.  However, beneath the surface there is still an important difference between these two approaches that belies their similar signatures.  For the effective state formalism to be renormalizable, the propagator must be modified so that it is consistent with how we have defined the effective state.  For the more conventional symmetry-breaking operators we have been studying here, we have used the standard Feynman propagator and so the renormalization proceeds more or less conventionally.

For example, if we consider operators that are quartic in the field such as 
\begin{equation}
{\cal L}_{\rm NR}^{(4)} 
= {\textstyle{1\over 36}} \lambda_1 K\varphi^4
+ {\textstyle{1\over 6}} \lambda_2 \varphi^3{\cal D}\varphi , 
\label{dim4q}
\end{equation}
or in the Robertson-Walker frame, 
\begin{equation}
{\cal L}_{\rm NR}^{(4)} 
= {\lambda_1\over 12} {H\over M} \varphi^4
+ {\lambda_2\over 6} {1\over aM} \varphi^3 \bigl( - \vec\nabla\cdot\vec\nabla \bigr)^{1/2}\varphi , 
\label{dim4qRW}
\end{equation}
these operators will generate one-loop divergent corrections.  The infinite parts of these corrections can then be readily removed by including counterterms of the form
\begin{equation}
K\varphi^2,\quad \varphi{\cal D}\varphi .
\label{counterterms}
\end{equation}

\section{Conclusions}
\label{conclude}

One of our goals here was to learn whether and to what extent simple symmetry-breaking operators could reproduce any of the various signatures generated by short-distance, ``trans-Planckian'' structures in the state of the inflaton.  This structure arises when nature is assumed to have some new physical principle---a shortest length scale, a non-commutativity or a quantum deformation of the classical symmetries of space-time, among many other possibilities---that would cause the actual vacuum to differ substantially from the flat-space vacuum at extremely short intervals.  

Although many such ideas have been applied to the vacuum state, they can largely be distinguished by whether they are established on a space-like or a time-like surface.  Within the former class are the ``effective state'' treatments \cite{greens,effective,schalm}.  One their more distinctive signatures, a correction to the power spectrum scaling as
\begin{equation}
{k\over k_*} \cos \biggl( 2 {k\over k_*} {M\over H} \biggr) , 
\label{standardA}
\end{equation}
can be exactly reproduced by a particular symmetry-breaking operator, described in Sec.~III.B.4.  Note that in this work, since we have assumed a standard Bunch-Davies vacuum throughout, we have not needed to modify the propagator as in the effect state approach \cite{greens,effective,schalm} to keep it consistent with the trans-Planckian structures in the state.  This close agreement between the predictions of these two effective approaches provides a new insight into the physical meaning of the effective states examined in \cite{greens,effective} since we can now see what sorts of more conventional symmetry-breaking operators are needed to produce the same effects.

The other class of vacuum states, whose structure is modified in the trans-Planckian regime, defines its states along a time-like surface.  In practice, what happens is that new eigenmodes of a quantum field are constantly being created to replenish earlier modes which have already red-shifted to longer wavelengths.  Each mode is first defined at a time $\eta_k$ when its wave number is equal to a cut-off scale, $M$,
\begin{equation}
{k\over a(\eta_k)} = M. 
\label{crossing}
\end{equation}
Since all of the modes are defined in exactly the same way, the ``ringing'' frequency in these models does not depend explicitly on $k$.  Instead, it depends solely on the natural time-evolution scale of the background, $H$, in addition to $M$.  Therefore, the typical correction to the power spectrum of this class, 
\begin{equation}
{H\over M} \cos \biggl( 2 {M\over H} \biggr) , 
\label{standardB}
\end{equation}
cannot be so readily mimicked by irrelevant symmetry-breaking operators, at least not the simple set that we have considered here.  In fact, such a signal might not ever be very naturally reproduced in an effective theory setting since a time-like boundary condition, such as these models apply, violates the basic assumption of causality which underlies any effective theory treatment.

Most of the work so far on testing Lorentz invariance has understandably concentrated on the possible signals of symmetry-breaking effects in high energy theory experiments \cite{relevant}.  Since the distances accessible to an accelerator experiment are extremely large, at least in comparison to the Planck scale, the experimentally important operators are the relevant or marginal ones.  Given a particular preferred frame---for example, one such as the spatially symmetric background that we studied here---it is not too difficult to determine all of the allowed operators in the Standard Model which are consistent with this symmetry \cite{sidney}.  

In an inflationary setting, the irrelevant symmetry-breaking operators can also produce measurable effects, at least in principle, since the dramatically rapid expansion of the universe effectively stretches short-distance structures to extremely large scales.  Although our emphasis here has been on these operators, we should note that the constraints on relevant operators, such as $K\varphi^2$ and $\varphi{\cal D}\varphi$, are even more stringent, since their relative contribution to the power spectrum scales as $M/H$.  Since very general irrelevant symmetry-breaking operators, such as those mentioned at the end of Sec.~IV, tend to require relevant operators for their renormalization, some fine-tuning of the parameters is inevitable if we are to keep small the contribution from these lower dimension operators in the renormalized theory.  However, our interest is primarily to compare with models with non-adiabatic vacuum structures---and moreover inflation is already plagued with many fine-tunings---so we have not much examined the question of naturalness here.

Tests of local Lorentz invariance provide insights into the structure of space-time at the tiniest scales.  Such tests continue to be important since there seems to be a basic incompatibility between the tenets of quantum field theory and those of general relativity at distances smaller than the Planck length.  Because of this impasse, it is especially vital to have some experimental guidance as to which of the postulates behind these two approaches ought to be preserved when formulating a yet more fundamental, inclusive theory of nature.

\begin{acknowledgments}

\noindent
This work was supported in part by DOE grant No.~DE-FG03-91-ER40682 and the National Science Foundation grant No.~PHY02-44801.  

\end{acknowledgments}

\appendix
\section{Integrals}

This short appendix describes some of the asymptotic behavior of the dimensionless integrals that we encountered when we calculated the corrections to the power spectrum from the symmetry-breaking operators.  Recall that the general form of these integrals is
\begin{eqnarray}
{\cal I}_n(z,z_0)
&\!\!\!=\!\!\!& \int_{z_0}^z {dz'\over z^{\prime n}}\, \Bigl[ 
\bigl[ 1 - z^2 + 4zz' - z^{\prime 2} + z^2z^{\prime 2} \bigr] \sin\bigl[ 2(z-z')\bigr] 
\nonumber \\
&&\qquad\quad 
-\, 2(z-z')[1+zz'] \cos\bigl[ 2(z-z')\bigr] \Bigr] . 
\label{Iintdefenc}
\end{eqnarray}
The specific cases that occur for the dimension-five operators are $n=1, 2, 3,$ and $4$.  Remember also that the arguments correspond to the final and initial conformal times scaled in terms of the wave number of a mode, $z=k\eta$ and $z_0=k\eta_0$.  

Inflation works by stretching a mode, which will eventually produce some observable feature of our universe, well outside of the Hubble horizon during inflation, 
\begin{equation}
{k\over |a(\eta)|} \ll H(\eta) , 
\label{latelimit}
\end{equation}
which for a de Sitter background becomes, 
\begin{equation}
k|\eta| \ll 1
\quad{\rm or}\quad
z = k\eta \to 0 .
\label{dSlatelimit}
\end{equation}
We shall therefore expand each of the relevant cases in this limit, 
\begin{eqnarray}
{\cal I}_4(z,z_0) 
&\!\!\!=\!\!\!& {2\over 3} {\rm Ci}(2z_0) 
- {(1+z_0^2)\sin(2z_0) - 2z_0\cos(2z_0)\over 3z_0^3} 
\nonumber \\
&&
- {2\over 3} \bigl[ \ln|2z| - 2 + \gamma \bigr] 
+ {\cal O}(z^2) 
\nonumber \\
{\cal I}_3(z,z_0) 
&\!\!\!=\!\!\!& {\rm Si}(2z_0) + {\cos(2z_0)\over z_0} 
- {1\over 2} {\sin(2z_0)\over z_0^2} 
+ {\cal O}(z^2) 
\nonumber \\
{\cal I}_2(z,z_0) 
&\!\!\!=\!\!\!& {3\over 2} + {1\over 2} \cos(2z_0) - {\sin(2z_0)\over z_0} 
+ {\cal O}(z^2) 
\nonumber \\
{\cal I}_1(z,z_0) 
&\!\!\!=\!\!\!& {1\over 2} z_0 \cos(2z_0) 
- {5\over 4} \sin(2z_0) + {\rm Si}(2z_0) 
+ {\cal O}(z^2) 
\nonumber \\
&&
\label{intsexpand}
\end{eqnarray}

The only absolute limit on the initial time is that---at the very least---relevant modes should have been within the Hubble horizon at the beginning of inflation, 
\begin{equation}
{k\over |a(\eta_0)|} > H(\eta_0) . 
\label{earlylimit}
\end{equation}
Again, in a de Sitter background this requires, 
\begin{equation}
z_0 = k\eta_0 < -1 .
\label{dSearlylimit}
\end{equation}
This bound is only saturated if inflation lasted just long enough that a mode of the size of the Hubble horizon is just re-entering the horizon today.  Most inflationary models produce much more expansion that this minimal amount and even in this extremal case most modes will be smaller than the Hubble horizon.  Therefore, we shall usually examine the limit $z_0\to -\infty$.  In any event, each of these integrals is of ${\cal O}(1)$ for $z_0=-1$ and $z$ small.

As we allow $\eta_0$ to extend arbitrarily far back in the past, we very soon encounter trans-Planckian modes---modes whose wavelength was smaller than a Planck length $1/M_{\rm pl}$ at the beginning of inflation, 
\begin{equation}
|z_0| > {M_{\rm pl}\over H} . 
\label{tPlimit}
\end{equation}
Let us therefore expand the integrals in the limit $z_0\to -\infty$ to learn how sensitively they can depend on these trans-Planckian modes, 
\begin{eqnarray}
{\cal I}_4(z,z_0) 
&\!\!\!=\!\!\!& - {2\over 3} \bigl[ \ln|2z| - 2 + \gamma \bigr]
+ {\cal O}(z^2,z_0^{-2}) 
\nonumber \\
{\cal I}_3(z,z_0) 
&\!\!\!=\!\!\!& 
{\pi\over 2}
+ {\cos(2z_0)\over 2z_0} 
+ {\cal O}(z^2,z_0^{-2}) 
\nonumber \\
{\cal I}_2(z,z_0) 
&\!\!\!=\!\!\!& {3\over 2} + {1\over 2} \cos(2z_0) 
+ {\cal O}(z^2,z_0^{-1}) 
\nonumber \\
{\cal I}_1(x,x_0) 
&\!\!\!=\!\!\!& {z_0\over 2} \cos(2z_0) + {\pi\over 2} 
- {5\over 4} \sin(2z_0)
+ {\cal O}(z^2,z_0^{-1}) . \qquad
\label{intbothexpand}
\end{eqnarray}
Of these four cases, it is the last that depends most sensitively on trans-Planckian physics, since it scales linearly with $z_0=k\eta_0$.

\end{document}